\documentclass[
preprintnumbers,
preprint,
a4paper,
showpacs, 
amsmath, 
amssymb, 
floatfix,
nofootinbib,
]{revtex4}

\usepackage[dvips]{graphicx}

\begin{document}

\title{Highly distorted apparent horizons and the hoop conjecture}

\author{Hirotaka Yoshino}

\affiliation{Department of Physics, University of Alberta, 
Edmonton, Alberta, Canada T6G 2G7}

\preprint{Alberta-Thy-22-07}

\date{December 23, 2007}

%
%
\begin{abstract}
By analyzing the apparent horizon (AH) formation in 
the collision of two {\it pp}-waves with rectangular sources
in four dimensions,
we study to what extent the AH can be distorted
without violating the energy conditions. 
It is shown that the highly distorted AH can form in this system
although it cannot be arbitrarily long.
The hoop conjecture is examined for the formation
of such highly distorted AHs, and 
our result gives a strong support to the hoop conjecture.
We also point out the possible relation between the AH topology
theorem and the hoop conjecture.
\end{abstract}

\maketitle

In $D$-dimensional spacetimes, 
the apparent horizon (AH) is defined as a $(D-2)$-dimensional surface
whose outgoing null geodesic congruence has zero expansion. The formation
of the AH implies the existence of the event horizon (EH) outside of it
if the null energy condition is satisfied \cite{W84}.
Although the black hole is usually defined by the EH,
the AH is also of interest since
the AH is a good indicator for the black hole formation.

The purpose of this paper is to examine to what extent the AH 
can be distorted in four-dimensional spacetimes.
In higher-dimensional spacetimes, it is known that
the AH of the spherical topology can be highly distorted.
For example,  in the system of the spindle-shaped matter in a four-dimensional
conformally-flat initial data, 
the AH can be arbitrarily long in some direction~\cite{IN02}. 
The distortion of four-dimensional static black holes
was also studied \cite{GH82}. Recently, it was shown that 
in the presence of a negative cosmological constant, 
the static black hole can be highly distorted in four dimensions \cite{T05}. 
In this paper, using a specific example, we study
whether the highly distorted AHs can form in dynamical situations 
in four-dimensional spacetimes without violating the energy conditions.
We give the system in which highly distorted AHs can actually
form, and discuss whether the formation of such highly distorted AHs is 
consistent with the hoop conjecture.

The hoop conjecture was proposed as an attempt to give the necessary and 
sufficient condition for the black hole formation.
{\it Black holes with horizons form when and only when 
a mass $M$ gets compacted into a region
whose circumference in every direction is
$\mathcal{C}\lesssim 4\pi GM$} \cite{Th72}.
The value $4\pi GM$ comes from the circumference $2\pi r_h(M)$ of 
the Schwarzschild black hole of mass $M$,  where $r_h(M):=2GM$
is the Schwarzschild radius.
This statement is often rephrased like ``the concentration of a mass
in every direction is necessary and sufficient for the black hole formation.''
The hoop conjecture has been tested using several systems, and no clear
couter-example was found so far.
Since this conjecture is loosely formulated, there are several discussions
on the definitions of the black hole (EH or AH),
the circumference $\mathcal{C}$, and the mass $M$ \cite{F91, CNNS94, YNT02}
(see also \cite{S07} for the recent trial to reformulate the hoop conjecture).
The meaning of ``$\lesssim$'' is also unclear. When the hoop conjecture
is tested using some systems, the authors usually choose the plausible
scale of the system as $\mathcal{C}$ and the ADM mass or some quasi-local
mass as $M$. Then they discuss whether $\mathcal{C}/2\pi r_h(M)$
becomes a parameter which indicates the AH formation
\cite{CNNS94, YNT02, NST88, W90, ST91, BIL91, T92, BGLL92, AHST92, C99}.
In this paper, we follow this direction and use the total energy of the system
as the definition of the mass $M$.

%
\begin{figure}[tb]
\centering
{
\includegraphics[width=0.3\textwidth]{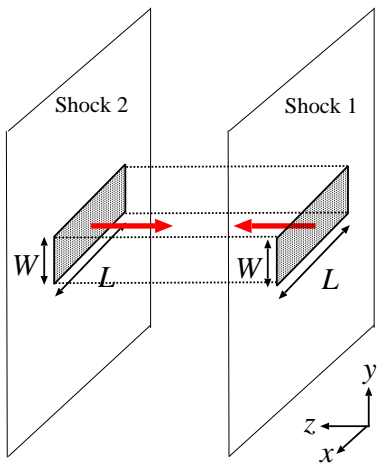}
}
\caption{Setup of the system. Two shock waves with rectangular sources
collide in four dimensions. The positions of the rectangles coincide at the
instant of the collision. The sizes of the rectangles in $x$ and $y$ directions
are $L$ and $W$, respectively.} 
\label{setup}
\end{figure}
%

Figure~\ref{setup} shows the system that we study in this paper.
Two shock waves propagate in $\pm z$ directions at the speed of light
and collide in a four-dimensional spacetime.
Each wave has the energy $p$ and there is
a rectangular source at the center. The  
sizes of the rectangle in $x$ and $y$ directions are $L$ and $W$, respectively. 
The positions of the two sources coincide when the shocks collide.
Since the waves are infinitely contracted, the energy
of the system is infinitely concentrated
in the $z$ direction at the instant of the collision.
The values of $L$ and $W$ indicate the degree of the concentration
of the energy in the  $x$ and $y$ directions, respectively.
We will show that the AH can be highly distorted for some values of $W$ and $L$.
However, it will be shown that the hoop conjecture holds also for
the formation of such highly distorted
AHs since the AH cannot be arbitrarily long in some direction.  
In the following, we adopt the gravitational radius
of the system energy $r_h(2p)$ as the length unit, i.e. $r(2p)=4Gp=1$.

In order to set up a shock wave with a rectangular source, 
we adopt the {\it pp}-wave metric
%
\begin{equation}
ds^2=-d\bar{u}d\bar{v}+d\bar{x}^2+d\bar{y}^2
+\Phi(\bar{x}, \bar{y})\delta(\bar{u})d\bar{u}^2,
\end{equation}
%
where $\delta(\bar{u})$ denotes the delta function.
The energy-momentum tensor of this spacetime
has the only nonzero component
$T_{\bar{u}\bar{u}}=\hat{\rho}(\bar{x}, \bar{y})\delta(\bar{u})$.
Here, $\hat{\rho}(\bar{x}, \bar{y})$ and $\Phi(\bar{x}, \bar{y})$
are related as
%
\begin{equation}
\bar{\nabla}^2\Phi=-16\pi G\hat{\rho}
\end{equation}
%
through the Einstein equation, where $\bar{\nabla}^2$
is the flat space Laplacian of the $(\bar{x}, \bar{y})$-plane.
We give the shock energy density $\hat{\rho}$ as
%
\begin{equation}
\hat{\rho}=p\vartheta_L(\bar{x})\vartheta_W(\bar{y}).
\label{energydensity}
\end{equation}
%
Here, $\vartheta_L(\bar{x})$ is defined by 
%
\begin{equation}
\vartheta_L(\bar{x}):=\frac{1}{L}\left[\theta(\bar{x}+L/2)-\theta(\bar{x}-L/2)\right],
\end{equation}
%
using the Heaviside step function $\theta(\bar{x})$. 
Then, the shock potential $\Phi$ is given by
%
\begin{equation}
\Phi=-\frac{1}{LW}\int_{-W/2}^{W/2}\int_{-L/2}^{L/2}
\log\left[(\bar{x}-\bar{x}^\prime)^2+(\bar{y}-\bar{y}^\prime)^2\right]
d\bar{x}^\prime d\bar{y}^\prime,
\end{equation}
%
in the unit $r_h(2p)=1$.
This is integrated as
%
\begin{equation}
\Phi=3-\frac{1}{LW}\sum_{\sigma_1, \sigma_2=\pm1}
\sigma_1\sigma_2
\left[
x_{\sigma_1}y_{\sigma_2}\log\left(\bar{x}_{\sigma_1}^2+\bar{y}_{\sigma_2}^2\right)
+\bar{x}_{\sigma_1}^2\arctan\frac{\bar{y}_{\sigma_2}}{\bar{x}_{\sigma_1}}
+\bar{y}_{\sigma_2}^2\arctan\frac{\bar{x}_{\sigma_1}}{\bar{y}_{\sigma_2}}
\right],
\end{equation}
%
where $\bar{x}_{\sigma}:=\bar{x}+\sigma L/2$ 
and $\bar{y}_{\sigma}:=\bar{y}+\sigma W/2$.
This metric reduces to that of the Aichelburg-Sexl particle \cite{AS71}
in the limit $W\to 0$ and $L\to 0$.

Since the coordinates $(\bar{u}, \bar{v}, \bar{x}, \bar{y})$
are discontinuous at $\bar{u}=0$,  we need to introduce the
continuous and smooth coordinates $(u, v, x, y)$.
Setting $(\bar{x}_1, \bar{x}_2)=(\bar{x}, \bar{y})$,
such coordinates are introduced by
%
\begin{align}
\bar{u}&=u;\nonumber\\
\bar{v}&=v+\theta(u)\Phi+\frac14u\theta(u)(\nabla\Phi)^2;\\
\bar{x}_i&=x_i+\frac12u\theta(u)\nabla_i\Phi.\nonumber
\end{align}
%
By this coordinate transformation, the metric becomes
%
\begin{equation}
ds^2=-dudv+H_{ik}H_{jk}dx^idx^j,
\end{equation}
%
%
\begin{equation}
H_{ij}=\delta_{ij}+\frac12u\theta(u)\nabla_i\nabla_j\Phi.
\end{equation}
%

Using the continuous and smooth coordinates,
we can set up the system of two {\it pp}-waves
by just combining the two metrics
%
\begin{equation}
ds^2=-dudv+\left[H_{ik}^{(1)}H_{jk}^{(1)}+H_{ik}^{(2)}H_{jk}^{(2)}-\delta_{ij}\right]dx^idx^j,
\end{equation}
%
%
\begin{align}
H_{ij}^{(1)}&=\delta_{ij}+\frac12u\theta(u)\nabla_i\nabla_j\Phi;\\
H_{ij}^{(2)}&=\delta_{ij}+\frac12v\theta(v)\nabla_i\nabla_j\Phi,
\end{align}
%
since the incoming waves do not interact before the collision.
This metric can be applied except at the interaction region $u>0, v>0$.

We study the AH on the slice $v\le 0=u$ and $u\le 0=v$.
On this slice, the AH is a union of two surfaces
$S_1$: $v=-\varPsi(x,y)$ in $v\le 0=u$ and 
$S_2$: $u=-\varPsi(x,y)$ in $u\le 0=v$.
The surfaces $S_{1}$ and $S_2$ are connected
on a common boundary $B$ in $u=v=0$.
The equation and the boundary conditions for the AH
on this slice were derived in \cite{EG02}. 
The AH equation is $\nabla^2(\varPsi-\Phi)=0$,
and the boundary conditions are
 $\varPsi=0$ and 
$\left(\nabla\varPsi\right)^2=4$ on $B$.    
The numerical method for solving this problem was established in \cite{YN03}.
We used the grid numbers $(50\times 50)$ for radial and angular coordinates
in most cases. Since the numerical error grows up as the AH becomes distorted,
we increased the grid numbers up to $(400\times 400)$ appropriately.

%
\begin{figure}[tb]
\centering
{
\includegraphics[width=0.4\textwidth]{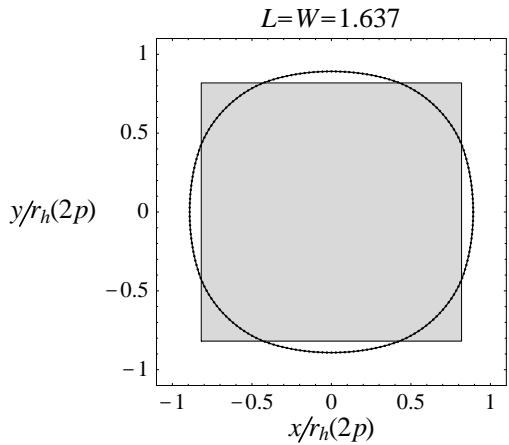}
}
\caption{The shape of the boundary $B$ (i.e., the cross section of the AH and $u=v=0$)
for $L=W=1.637$. The shape of the source is indicated by the grey region.
For $L=W>1.637$, we could not find the AH. The distortion parameter
$\zeta:=r_{\rm max}/r_{\rm min}$ is $1.05$.} 
\label{C-sameLW}
\end{figure}
%

Now we show the numerical results. In order to 
discuss the degree of distortion of the AH, 
we introduce the spherical-polar coordinates $(r, \theta)$ 
in the $(x,y)$-plane. In these coordinates, the boundary $B$
is given as $r=f(\theta)$. Then, the
parameter $\zeta:=r_{\rm max}/r_{\rm min}$
gives a good indicator for the degree of distortion,
where $r_{\rm max}$ and $r_{\rm min}$
are the maximum and minimum values of $f(\theta)$, respectively.
Let us first look at the case $L=W$, where the source has the shape of a regular square. 
In this case, the boundary $B$ becomes small and 
a little bit distorted as the value of $L(=W)$
is increased. We could not find the AH for $L=W>1.637$. 
Figure~\ref{C-sameLW} shows the shape of $B$
for $L=W=1.637$. 
For this value of $L(=W)$,  it is found that $r_{\rm min}=f(0)=f(\pi/2)$
and $r_{\rm max}=f(\pi/4)$, and the distortion parameter is $\zeta\simeq 1.05$.

%
\begin{figure}[tb]
\centering
{
\includegraphics[width=0.5\textwidth]{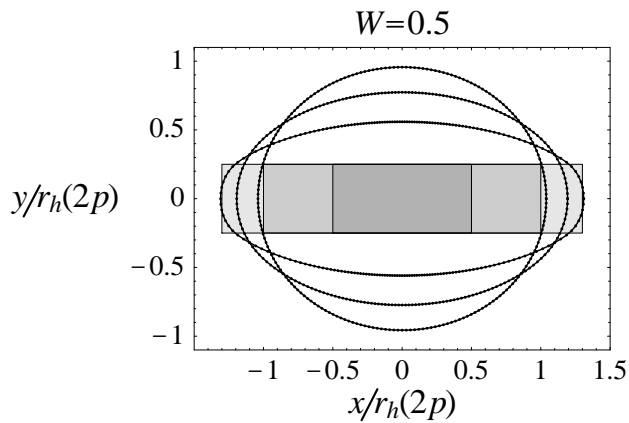}
}
\caption{The shapes of the boundary $B$ for $W=0.5$ and $L=1.00, 2.00, 2.60$. 
The shapes of the source are indicated by the grey regions.
For $L\ge 2.61$, we could not find the AH. For $L=2.60$,
the distortion parameter is $\zeta=2.34$.} 
\label{C-shape-W0.5}
\end{figure}
%
%
\begin{figure}[tb]
\centering
{
\includegraphics[width=0.5\textwidth]{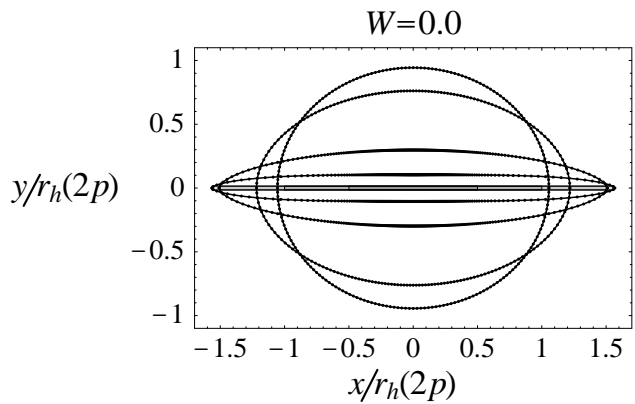}
}
\caption{The shapes of the boundary $B$ for $W=0.0$ and $L=1.00, 2.00, 3.00, 3.13$. 
The shapes of the source are indicated by the grey regions.
For $L\ge 3.14$, we could not find the AH. For $L=3.13$,
the distortion parameter is $\zeta\simeq 15$ and the AH is highly distorted.} 
\label{C-shape-W0.0}
\end{figure}
%

Next, we fix $W$ and increase $L$. Let us choose $W=0.5$
as an example. The shapes of the boundary $B$ for $L=1.00, 2.00, 2.60$ are shown 
in Fig.~\ref{C-shape-W0.5}. We could not find the solution of the AH for $L\ge 2.61$.
For these values of $W$ and $L$, $r_{\rm min}=f(\pi/2)$ and $r_{\rm max}=f(0)$.
The distortion parameter $\zeta$ becomes larger as $L$ is increased,
and it is $\zeta\simeq 2.34$ for $L=2.60$.

Let us look at the case $W=0.0$, where we can find the highly
distorted AH. Figure~\ref{C-shape-W0.0}
shows the shapes of the boundary $B$ for $L=1.00, 2.00, 3.00, 3.13$.
We could not find the AH for $L \ge 3.14$.
As we can see, the AH is highly distorted for $L = 3.13$. 
The distortion parameter $\zeta$ has the tendency to become
larger as $W$ is decreased and $L$ is increased for $L>W$,
and it takes the maximum value  for $W=0.0$ and $L=3.13$. 
This maximum value of $\zeta$ is more than $15$
and much larger than unity. Therefore, the highly distorted AHs can form
in this system. But we point out that the value of
the maximum radius $r_{\rm max}$ of the boundary $B$
is restricted from above as $r_{\rm max}\lesssim 1.57$.
Therefore, the AH cannot become arbitrarily long in some direction.

%
\begin{figure}[tb]
\centering
{
\includegraphics[width=0.5\textwidth]{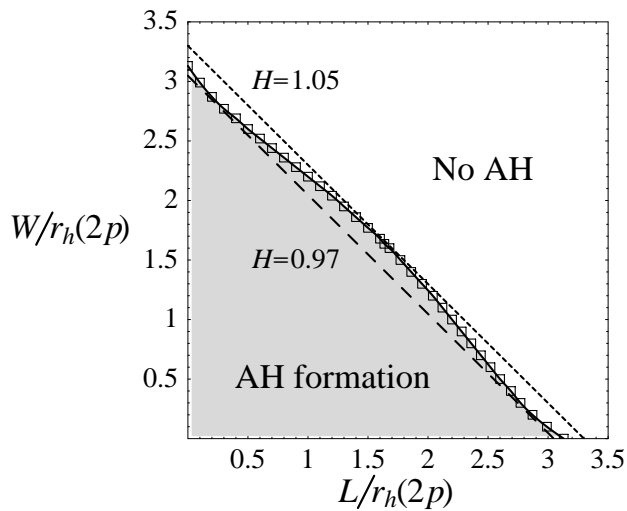}
}
\caption{The region of the AH formation in the $(L,W)$-plane (the grey region). 
The border of this region is shown by a solid line, on which
the numerical data is shown by squares ($\square$). The contours of $H=0.97$
and $H=1.05$ are shown by a dashed line and a dotted line, respectively.
The AH forms if $H\le 0.97$ and does not form if $H\ge 1.05$.  
} 
\label{AHcondition}
\end{figure}
%

We test the hoop conjecture using our result. 
For this purpose, it is convenient to introduce a parameter 
\begin{equation}
H:=\frac{\mathcal{C}}{2\pi r_h(M)}.
\end{equation}
In this paper, we adopt the
circumference $2(W+L)$ of the source as the definition of 
$\mathcal{C}$ and the total energy $2p$ of the system 
as the definition of $M$. The grey region in Fig.~\ref{AHcondition} 
shows the region of the AH formation 
in the $(L,W)$-plane.
The contours of $H=0.97$ and $H=1.05$ are also shown.
From this figure, we see that the AH forms if $H\le 0.97$ and
the AH does not form if $H\ge 1.05$.
Therefore, the parameter $H$ gives a good indicator for the AH formation
in this system and our result is consistent with the hoop conjecture. 
The hoop conjecture holds well also for the formation
of the highly distorted AHs. This is because the AH can be highly distorted,
but cannot be arbitrarily long in some direction.

It is worth pointing out that the highly distorted AH can form
also in the system of collapsing convex null-dust shell \cite{BIL91}. 
For a cylindrical shell with two hemisphere caps, 
the AH can form for very small radius of the cylinder $r\ll r_h(M)$,
while the length of the cylinder can be as large as $4r_h(M)$
 (see Fig.1(b) in \cite{BIL91}).
Therefore, it is expected that in many systems the highly distorted AHs 
can form but its size in the largest direction
is restricted from above.

Let us discuss whether the formation of a highly distorted AH
leads to an interesting gravitational phenomena. 
In the higher-dimensional cases, one would expect that
the Gregory-Laflamme instability \cite{GL87} occurs when a highly 
distorted AH forms. However, 
such an instability is not known in the four-dimensional case, since
there is no black string solution.
Although there is a four-dimensional cylindrical black hole solution
in the presence of a negative cosmological constant $\Lambda$,
it turns out to be stable \cite{CL01}. This is in contrast to the fact that
the uniform AdS black strings in higher dimensions
are unstable for sufficiently small $|\Lambda|$ \cite{BDR07}.
Therefore, we cannot expect the interesting phenomena
such as the pinch-off of the AH. 
A highly distorted AH could form in a dynamical situation,
and it will become less distorted in the temporal evolution.
In our system, the final state is expected to be a Schwarzschild black hole.

The mass of the final Schwarzschild black hole $M_{\rm BH}$ is determined
by the amount of gravitational radiation. Using the area theorem,
we can evaluate the lower bound on the mass of the final state
as 
\begin{equation}
M_{\rm AH}=\frac{1}{G}\sqrt{\frac{A_{\rm AH}}{16\pi}},
\end{equation}
with the AH area $A_{\rm AH}$. In our system, there is a tendency
that $M_{\rm AH}$ becomes smaller as the AH becomes more distorted.
In the case $W=0.0$ and $L=3.13$,
$M_{\rm AH}$ is less than 30\% of the total system energy $2p$. 
Although $M_{\rm AH}$ is just the
lower bound on $M_{\rm BH}$, it is natural to expect that there is
some correlation between $M_{\rm AH}$ and $M_{\rm BH}$.
Therefore, if a highly distorted AH forms in some system,
a lot of gravitational wave could be radiated.

Finally, we point out the possible relation between the theorem
on the AH topology \cite{H72} and the hoop conjecture.
For this purpose, let us recall the study of \cite{IN02} on  
the AH formation in several systems in the momentarily-static
conformally-flat four-dimensional space 
(i.e., the initial data of the five-dimensional spacetime). 
They showed that (i) the arbitrarily long AH of the spherical topology
$S^3$ forms for the spindle-shaped 
matter distribution and (ii) the AH of the ring topology $S^1\times S^2$ forms 
for the ring-shaped matter distribution, if the ring radius is sufficiently large. 
The result of (ii) is naturally expected
from the result of (i), since the ring-shaped matter distribution
is achieved by bending the spindle-shaped matter distribution.
Therefore it is indicated that if a long AH of the spherical topology can form, 
an AH of the ring topology also can form. Conversely, it is expected that
if an AH of the ring topology cannot form, a long AH of the spherical topology
cannot form. Since an AH of the ring topology 
(or equivalently the torus topology $S^1\times S^1$) 
is forbidden in four dimensions \cite{H72}, the formation of a long AH is also
expected to be prohibited. This statement is consistent with the
``only when'' part of the hoop conjecture.

The above discussion gives one plausible interpretation for 
the reason why the ``only when'' part 
of the hoop conjecture holds in four dimensions.
We also hope that this observation
could give a hint for studies which attempt to prove 
the ``only when'' part of the hoop conjecture,
since at least physically the AH topology theorem is related  
to the hoop conjecture. Here, we would like to
note that currently there is no theorem which corresponds to the ``only when''
part of the hoop conjecture, though there is a strong theorem
by Schoen and Yau \cite{SY83} 
which corresponds to the ``when'' part of the hoop conjecture.

To summarize, we studied the collision of shock waves with 
rectangular sources, and found that the highly distorted
AH can form in this system although it cannot be arbitrarily long.  
The hoop conjecture remarkably holds also for the formation
of such highly distorted AHs, 
and our result gives a strong support to the hoop conjecture. 
The higher-dimensional generalization
of the study in this paper, especially in the case $W=0$, is interesting in the context
of the black hole production at accelerators in TeV gravity scenarios. 
It will be reported in our forthcoming paper.

The author thanks the Killam trust for financial support.



\end{document}